\documentclass[a4paper,11pt]{article}
\usepackage{pos}
\usepackage{comment}
\usepackage{hyperref}
\newcommand{\orcid}[1]{\href{https://orcid.org/#1}{\includegraphics[width=8pt]{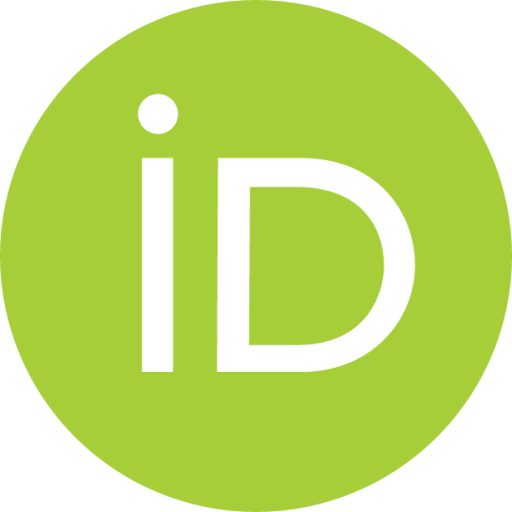}}}

\title{Searching for Ultralight Dark Matter Solitons with Gravitational Waves}

\author{Silvia Gasparotto \orcid{0000-0001-7586-1786}}

\affiliation{Grup de F\'{i}sica Te\`{o}rica, Departament de F\'{i}sica, Universitat Aut\`{o}noma de Barcelona, 08193 Bellaterra (Barcelona), Spain}
\affiliation{Institut de F\'isica d'Altes Energies (IFAE), The Barcelona Institute of
Science and Technology (BIST), Campus UAB, 08193 Bellaterra, Barcelona}

\emailAdd{sgasparotto@ifae.es}

\abstract{
Ultralight Dark Matter (ULDM) offers an alternative to cold dark matter, characterized by wavelike behavior on galactic scales. This contribution summarizes our work~\cite{Blas:2024duy} on how solitonic cores induced by ULDM affect gravitational waves (GWs), enabling their detection through next-generation observatories like LISA, Einstein Telescope and Cosmic Explorer. We show that continuous GWs from spinning neutron stars near the Galactic Center (GC) can probe solitons for \(m \lesssim 10^{-22}\, \mathrm{eV}\) and surpass current constraints on the ULDM couplings to the Standard Model for masses  \(m \lesssim 10^{-20}\, \mathrm{eV}\). Additionally, GW modulation could reveal solitons in extragalactic systems, particularly in more massive halos, highlighting the broad potential of GW-based methods for uncovering ULDM properties.}
\begin{document}
\maketitle

\section{Introduction}

Elucidating the nature of dark matter (DM) remains a key challenge in modern physics. Ultralight Dark Matter (ULDM) offers an interesting framework, characterized by a large de Broglie wavelength,  
\[
\lambda_{\text{db}} \sim 0.5 \, \text{kpc} \left( \frac{10^{-22} \, \text{eV}}{m} \right) \left( \frac{250 \, \text{km} \, \text{s}^{-1}}{v} \right),
\]  
which gives rise to wavelike behavior on galactic scales. DM particles with masses smaller than \(\sim 1 \, \mathrm{eV}\)~\cite{Hui:2021tkt} can indeed be effectively described by \emph{classical wave} equations for a massive scalar field \(\phi\) interacting through gravity (and potentially additional interactions).  

This wavelike nature results in two primary consequences. First, ULDM exhibits unique dynamics on scales smaller than \(\lambda_{\rm db}\), such as the formation of solitons—dense, self-gravitating structures located at the centers of galaxies—surrounded by Navarro-Frenk-White (NFW) profiles at larger radii~\cite{Schive:2014dra}. Second, coherent oscillations of the ULDM field occur at a frequency determined by its mass, \(\omega \sim m \approx \frac{m_{22}}{76 \, \mathrm{days}}\) where we define $m_{22}=m/10^{-22}$ eV. The solitonic cores can be constrained by galactic rotation curves~\cite{Bar:2018acw}, which disfavor ULDM models with masses \(m \lesssim 10^{-21} \, \mathrm{eV}\). Additionally, time-dependent oscillations of the gravitational potential produce detectable patterns in Pulsar Timing Arrays (PTAs)~\cite{Khmelnitsky:2013lxt}, although probing the Galactic Center (GC) with pulsar data remains challenging due to interstellar medium effects.  

This contribution, which summarises our work~\cite{Blas:2024duy}, extends the study of ULDM-induced oscillations to gravitational waves (GWs), which are less subject to photometric constraints and allow the exploration of higher ULDM masses. We show how GWs emitted by spinning neutron stars (NSs) will provide an independent probe of the soliton at the centre of our Galaxy for masses \(m < 10^{-22} \, \mathrm{eV}\). Furthermore, this method will provide a way to explore regions of parameter space at higher masses that are currently loosely constrained, in particular, in scenarios involving direct couplings between ULDM and Standard Model fields.  

This work is structured as follows. Section~\ref{sec2} introduces the essential formalism for oscillating gravitational potentials and redshifts. Section~\ref{sec3} examines their effects on GWs and discusses detectability. Section~\ref{sec4} explores the implications for the Milky Way, followed by concluding remarks and possibile future directions in Section~\ref{sec5}.

\section{ULDM phenomenology: soliton and metric oscillations}
\label{sec2}

ULDM, modeled as a scalar field \(\phi\), satisfies the Klein-Gordon equation \(\Box \phi + m^2 \phi = 0\), which resembles the harmonic oscillator equation, where the scalar field's mass sets the frequency, \(\omega = m\). Within the galaxy, the spatial distribution of the ULDM field is of primary interest. The field can be parameterized as  
\[
\phi \equiv \frac{\psi \, e^{-i m t}}{\sqrt{2 m}} + \text{c.c.},  
\]  
where rapid oscillations are factored out. The envelope function \(\psi\) evolves on timescales much longer than the coherent oscillation period \(1/m\), depending on its self-energy scale \(\gamma \ll m\) neglected here.  
In the non-relativistic regime, and adopting the metric  
\[
{\rm d}s^2 \approx -(1-2\Phi)\,{\rm d}t^2 + (1+2\Psi)\delta_{ij}\,{\rm d}x^i\,{\rm d}x^j,  
\]
the Klein-Gordon equation reduces to the Schrödinger-Poisson system:  
\begin{subequations} \label{eqs:SP}%
\begin{align}
    i \partial_t \psi &= -\left(\frac{\nabla^2}{2m} + m \Phi \right)\psi, \label{eq:S}  \\
    \nabla^2 \Phi &= -4 \pi (\rho_\phi + \rho_{\rm m}), \quad \Psi \approx \Phi, \label{eq:P}
\end{align}
\end{subequations}  
where \(\nabla^2 \equiv \delta^{ij} \partial_i \partial_j\), \(\rho_\phi \equiv m |\psi|^2\) is the ULDM density, and \(\rho_{\rm m} \equiv T^{\rm m}_{\;tt}\) is the matter density.  
Numerical simulations of these equations consistently reveal the formation of a dense, coherent structure—known as \textit{soliton}—at the galactic center, emerging from the condensation of the DM halo. The soliton's density profile can be computed as a stationary, spherically symmetric ground-state solution of Eqs.~\eqref{eqs:SP}, given by:  
\begin{equation}
\label{eq:rhosol}
\rho_{\text{sol}} = \frac{\rho_0}{\left(1 + 0.091 \left(\frac{r}{r_c}\right)^2\right)^8}, \quad \text{with} \quad r_c \sim 0.4 \lambda_{\rm db},  
\end{equation}
where \(\rho_0\) is the central density. The total soliton mass relates to the DM halo mass via \(M_{\text{sol}} \approx 1.4 \times 10^9 \left({10^{-22} \, \text{eV}}/{m_{\text{dm}}}\right) \left({M_{\text{halo}}}/{10^{12} \, M_{\odot}}\right)^{1/3}\), although some variability exists across simulations~\cite{Chan:2021bja}.  

Equation \eqref{eq:rhosol} describes the stationary average density of the overdense region, while the scalar field's oscillatory nature introduces a fast, time-dependent component \(\delta\rho\). For a quadratic potential, \(\delta\rho\) oscillates at a frequency \(\omega = 2m\), with higher-order corrections (e.g., \(\omega = 4m\)) from self-interactions being neglected here (see Ref.~\cite{Blas:2024duy} for the case with also self-interactions). This oscillatory density arises from second-order terms in the energy-momentum tensor and depends on the field’s velocity \(v\) and self-energy \(\gamma\). Through the Einstein equations, \(\delta\rho\) generates a time-dependent gravitational potentials, \(\delta\Phi_2\) and \(\delta\Psi_2\), oscillating at the same frequency  \(\omega = 2m\). The dominant term  satisfies \(\nabla^2 \delta\Psi_2 = -4\pi \delta\rho\), with $\delta\rho = \nabla^2(\rho / 4m^2)$.  

Another oscillatory effect on the metric arises if ULDM directly interacts with Standard Model fields. A straightforward, yet phenomenologically rich option, is to consider a \emph{universal} conformal coupling of the ULDM field $\phi$ to ordinary matter. This interaction can be reformulated through a conformal transformation to the Jordan-Fierz frame with metric is $ \widetilde{g}_{\mu\nu} = A^2(\phi) g_{\mu\nu} $, for which test (ordinary) matter fields follow geodesics. For simplicity, we considered the interaction parametrized as $ A \approx 1 + \phi / \Lambda_1 $ (linear) or $ A \approx 1 + \phi^2 / \Lambda_2^2 $ (quadratic), with the high-energy scales $\Lambda_1$ and $\Lambda_2$ parameterizing the interaction strength. These couplings induce gravitational field oscillations with frequencies \(\omega=m\)
(linear) or \(\omega=2m\) (quadratic).

\section{Propagation of radiation in ULDM background: the case of GWs}
\label{sec3}

Inhomogeneities in the matter distribution modify the propagation of signals through their perturbation of space-time, primarily through the well-known \textit{gravitational redshift} of the signal frequency \(\omega_e\). In the geometric optics limit, where the signal’s frequency $\omega_e$ is much higher than the oscillation frequency of the background \(\omega_m\) (\(\omega_e \gg \omega_m\)), this redshift is given by:  
\begin{equation}\label{eq:relrecem}
    \frac{\Delta\omega_{\rm e}}{\omega_{\rm e}} \approx \Psi|^{\rm r}_{\rm e} + n^i v_i|^{\rm r}_{\rm e} - n^i \int_{\lambda_{\rm e}}^{\lambda_{\rm r}} \partial_i (\Phi + \Psi) \, {\rm d}\lambda',
\end{equation}  
where \(n^i\) is the unit vector along the path from emission to observation and, for instance, the potential $\Psi|^{\rm r}_{\rm e}$ is evaluated at emission and receiver position. Decomposing the gravitational potentials as \(\Psi = \Bar{\Psi} + \delta \Psi_2 \cos(2mt)\) (and similarly for \(\Phi\)), the time-independent term \(\Bar{\Psi}\) leads to a constant frequency shift, while the oscillatory term \(\delta \Psi_2\) induces a frequency modulation, leaving distinct features in the data.  

This effect, first studied for millisecond pulsars~\cite{Khmelnitsky:2013lxt}, constrains the ULDM density at the pulsars’ locations (approximately 1 kpc from Earth) for masses in the range \(10^{-24} \, \mathrm{eV} \lesssim m \lesssim 10^{-22} \, \mathrm{eV}\)~\cite{Smarra:2024kvv}. The same modulation affects the frequency of gravitational waves (GWs). Parameterizing the modulation as \(\Delta \omega_e / \omega_e = \Upsilon \cos(\omega_\delta t + \varphi_\delta)\), the parameter \(\Upsilon\) depends on the nature of the coupling:  
\begin{gather}
    \Upsilon \equiv 
    \begin{cases}
        \big[\Psi_2 - \frac{2}{\omega_\delta} n^i \partial_i \Phi_2 \big]_{x^i_{\rm e}}, & \text{(minimal coupling)} \\
        \frac{\sqrt{2}}{\Lambda_1} \left( \frac{\Bar{\rho}_\phi(x^i_{\rm e})}{m^2} \right)^{1/2}, & \text{(direct linear coupling)} \\
        \frac{1}{\Lambda_2^2} \frac{\Bar{\rho}_\phi(x^i_{\rm e})}{m^2}, & \text{(direct quadratic coupling)}.
    \end{cases} \label{eq:upps}
\end{gather}  
This modulation causes a monochromatic GW, \(h_{\text{GW}} = \mathcal{A} \cos(\omega_e t + \alpha_0)\), to develop sidebands:  
\begin{equation}\label{eq:hplusdelta}
    h \approx \mathcal{A} \Big[\cos(\omega_e t + \alpha_0') \pm \frac{\omega_{\rm e}}{2\omega_\delta} \Upsilon|_{x_{\rm e}} \cos[(\omega_e \pm \omega_\delta) t + \alpha_0' \pm \varphi_\delta]\Big],
\end{equation}  
where \(\alpha_0\) and \(\varphi_\delta\) are constant phases, and just terms of \(\mathcal{O}(\omega_{\rm e}/\omega_\delta \cdot \Upsilon)\) are retained. The sidebands at \(\omega_e \pm \omega_\delta\) are distinguishable from other phase modifications, such as post-Newtonian effects.

\begin{figure}[t!]
    \centering
    \includegraphics[scale=0.4]{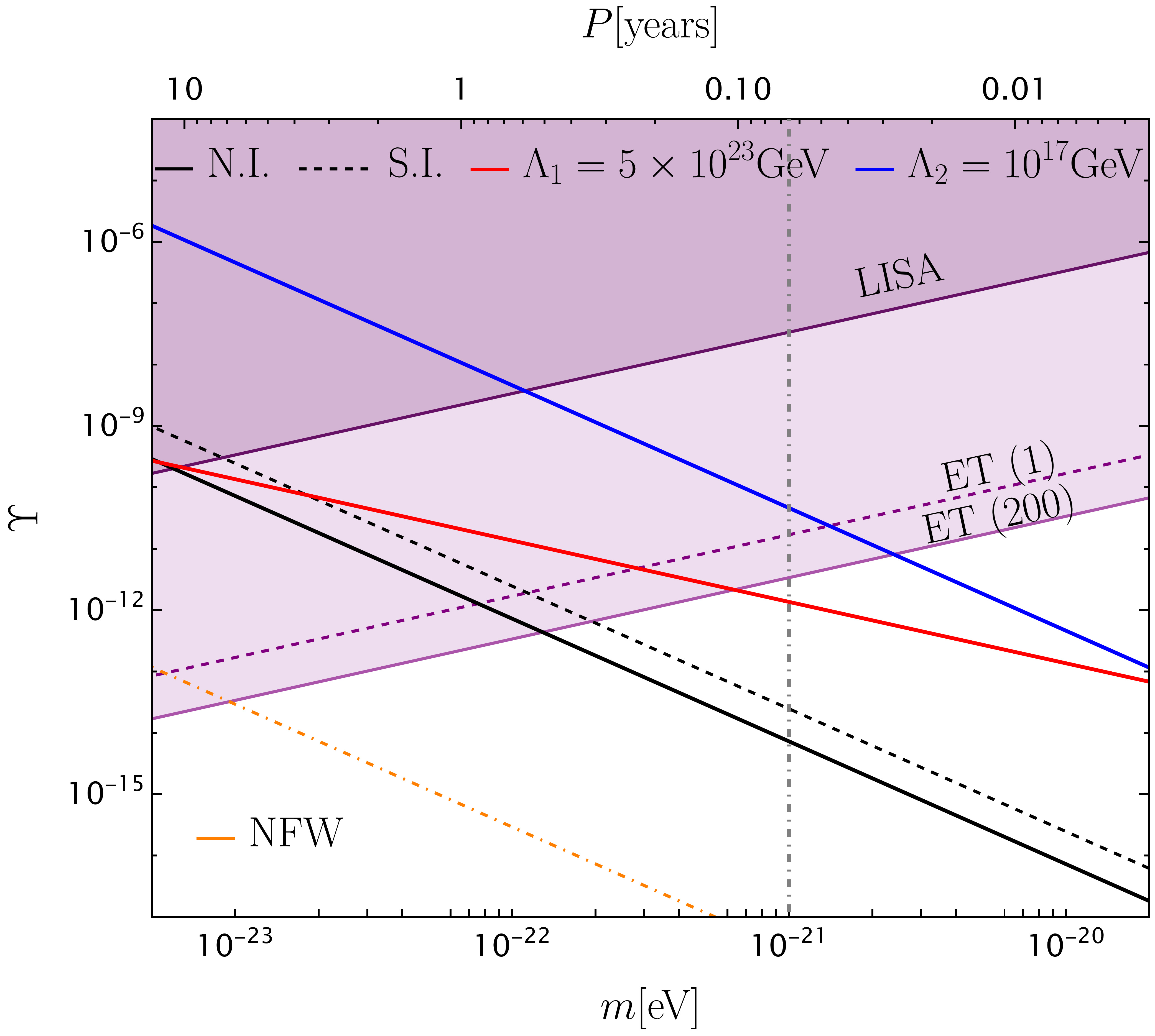}
    \caption{Sensitivities of LISA and ET/CE to~$\Upsilon$ (cf. Eq.~\eqref{eq:upps}) from Ref.~\cite{Blas:2024duy}. 
    The region in dark (light) purple is probed by LISA (ET/CE) with X-MRIs (spinning NS). 
    The dashed purple line instead represents the sensitivity for a single young NS with $f_{\rm e}=10^3\,{\rm Hz}$ and ${\rm SNR}=20$. The amplitudes of $\Upsilon$ are shown in black-solid for no self-interactions (N.I.) and black-dashed for critical self-interactions (S.I.) in the minimal coupling case. The blue (red) line corresponds to an example case of direct interaction with quadratic (linear) coupling. All curves correspond to a soliton at the GC with central density~$\bar{\rho}_0=10^3 M_\odot {\rm pc}^{-3}$, except the orange which is for the Galactic NFW density profile at $r_{\rm e}=500\, {\rm pc}$. ULDM with masses below the grey dashed line is disfavoured as the dominant component DM. The top horizontal axis shows the period corresponding to~$\omega =2 m$.}
    \label{fig:Upsilonres}
\end{figure}
\begin{figure}
    \centering
    \includegraphics[width = 0.49 \linewidth]{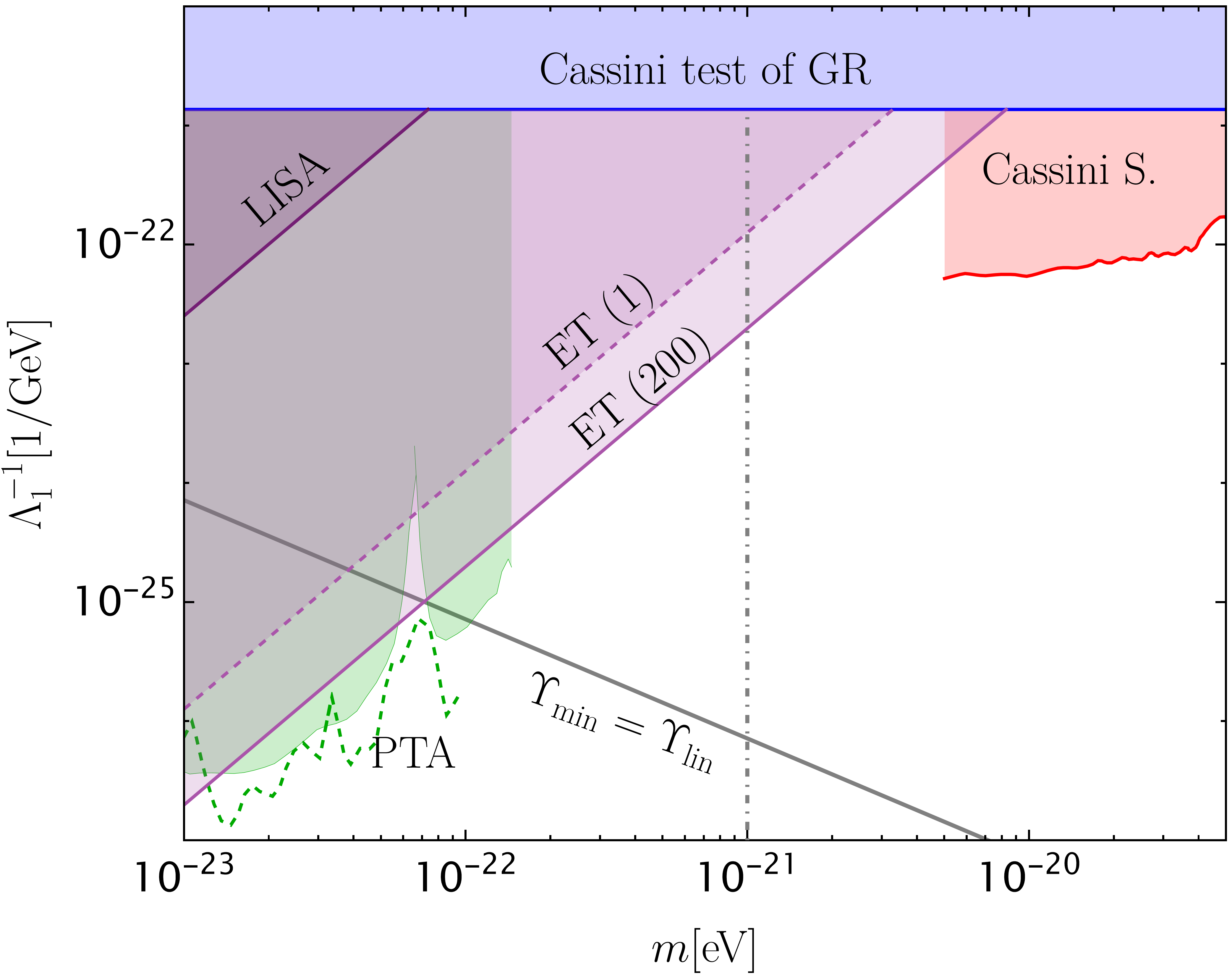}
    \includegraphics[width = 0.49 \linewidth]{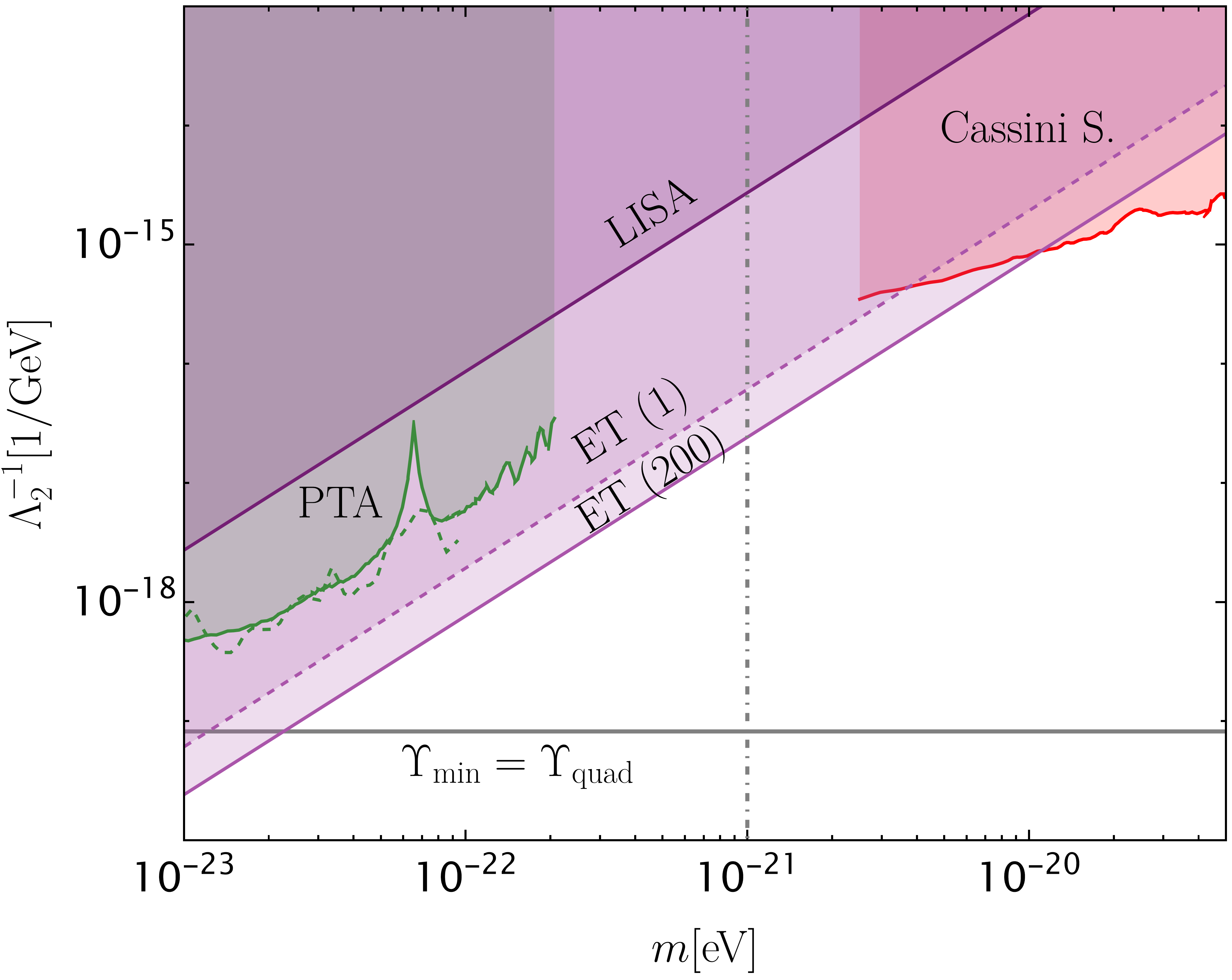}
    \caption{ From Ref.~\cite{Blas:2024duy},
    sensitivities of LISA and ET/CE to the linear direct coupling~$\Lambda_1^{-1}$ \textit{(left)} and quadratic coupling~$\Lambda_1^{-2}$ \textit{(right)} from the modulation of GWs by a ULDM soliton at the GC with central density~$\bar{\rho}_0=10^3 M_\odot {\rm pc}^{-3}$. The coloured regions are excluded by PTA (green), Cassini tests of General Relativity (blue), and Cassini bounds on a stochastic GW background (red). As in~Fig.~\ref{fig:Upsilonres}, in the case of ET/CE we show lines both for a single observation of a spinning NS and for a population of $\mathcal{O}(200)$. Here, we choose a threshold of SNR$_{\rm th}=5$ for a more fair comparison with the other bounds. The gray line represents values where the effect in the minimal case is stronger than that observed with direct interaction in our case (excluding PTA). Thus, any values below this line indicate a definite detection of the effect.
   } 
   \label{Fig:lambdares}
\end{figure}
\section{ULDM sensitivities from Galactic GW sources }
\label{sec4}
The modulation effects of ultralight dark matter (ULDM) are most pronounced in dense regions, such as galactic centers. However, photometric observations near the Galactic Center (GC) are often hindered by significant dust contamination, making it challenging to obtain clean data from sources like millisecond pulsars. Gravitational waves (GWs), unaffected by dust, offer a promising alternative for probing ULDM solitons in these dense regions, where baryonic density is also higher.
In the Milky Way, photometric data suggests that the average density within 1 kpc of the GC is approximately \(\rho \sim 10^3 \, M_\odot/\mathrm{pc}^3\), consistent with the presence of a soliton for a scalar mass \(m \sim 10^{-22} \, \mathrm{eV}\). Sensitivity to ULDM effects can be significantly enhanced by combining evidence from \(N\) GW sources. The total signal-to-noise ratio (SNR) for detecting modulation effects can be approximated as:  
\begin{equation}\label{eq:Gformula}
    \mathrm{SNR}_\delta \sim \frac{1}{\sqrt{2}} \left(\frac{\omega_{\rm e}}{\omega_\delta}\right) \Upsilon(\bar{\rho}_0, m, x_{\rm e}^i) \sqrt{N} \langle \mathrm{SNR}_{h} \rangle,
\end{equation}  
where \(\langle \mathrm{SNR}_{h} \rangle\) is the average SNR for a given GW population.

We analyzed three distinct populations of monochromatic GWs in the Milky Way: white dwarf binaries, extremely large mass-ratio inspirals (X-MRIs), and deformed spinning neutron stars. The first population, white dwarf binaries, emits in the mHz regime and is a guaranteed source of (quasi-)monochromatic GWs. Based on population models~\cite{Korol:2021pun}, we estimate approximately $\sim 10^3$ resolved binaries detectable by LISA within the first kpc from the GC, with an average SNR of $\sim 40$ (see Table I in Ref.~\cite{Blas:2024duy}). The second population consists of X-MRIs, which involve brown dwarfs orbiting Sgr A* with extremely large mass ratios. Predictions from~\cite{Amaro-Seoane:2019umn} suggest that a few such systems emitting in the mHz regime could be detected by LISA, with SNRs reaching $\sim 10^3$, then $\omega_e\sqrt{N} \langle \mathrm{SNR}_{h}\rangle\sim 10$. Despite their differences, both populations exhibit comparable sensitivities when evaluated using Eq.~\eqref{eq:Gformula}, and are represented with a single line in Figure~\ref{fig:Upsilonres}.  The third population comprises deformed spinning neutron stars, which emit continuous GWs at much higher frequencies (\(f \gtrsim 0.1 \, \mathrm{kHz}\)). Next-generation ground-based interferometers like the Einstein Telescope (ET) and Cosmic Explorer (CE) may detect several of these sources, potentially up to $\mathcal{O}(200)$ in the most optimistic scenarios~\cite{Pagliaro:2023bvi}. Since the SNR scales linearly with the carrier frequency, this population benefits from a large sensitivity gain due to its high frequencies, leading to $\omega_e\sqrt{N} \langle \mathrm{SNR}_{h}\rangle\sim 10^5$.

In Figure~\ref{fig:Upsilonres}, we have compared the sensitivity of LISA for X-MRIs with that of the Einstein Telescope (ET) for a single rotating neutron star (NS) characterised by \(f_e \sim 10^3 \, \mathrm{Hz}\) and \({\rm SNR}_h \sim 20\) (light purple), and for a population of \(\mathcal{O}(200)\) such neutron stars. The expected amplitudes of the soliton-induced oscillations are also shown as a solid black line for the non-interacting case and as a dashed line for the interacting case. Our analysis shows that the sources observed by ET outperform those detected by LISA, providing significantly higher sensitivity for probing soliton effects in the Galactic Centre. Remarkably, even a single (young) rotating neutron star near the GC could be sufficient to probe the ULDM soliton for scalar field masses \(m \lesssim 10^{-22} \, \mathrm{eV}\). However, a key feature of this scenario is that if a soliton exists, it should induce modulation effects in all GW sources originating from that region. This universality provides a robust method for confirming the presence of a soliton by cumulative evidence from an entire GW population.

In addition, observations with ET/CE would surpass LISA in probing direct couplings of ULDM to the Standard Model (cf. Figures~\ref{Fig:lambdares}). These detectors would not only cover the mass range \(2 \times 10^{-22} \lesssim m \, [\mathrm{eV}] \lesssim 3 \times 10^{-21}\) with unprecedented sensitivity for both linear and quadratic couplings, but also exceed the current PTA (Pulsar Timing Array) constraints on quadratic couplings for \(m \lesssim 2 \times 10^{-22} \, \mathrm{eV}\). For comparison, we also include constrained regions from the PTA~\cite{Smarra:2024kvv}, Cassini tests of general relativity~\cite{Blas:2016ddr,Bertotti:2003rm}, and Cassini bounds on the GW stochastic background~\cite{Blas:2016ddr,Armstrong:2003ay}.

\section{Conclusions and Future Work}
\label{sec5}
In this work, we demonstrated the potential of gravitational waves to probe ULDM through modulation effects induced by solitonic cores. Next-generation detectors such as the Einstein Telescope and Cosmic Explorer, which are likely to detect continuous GWs from spinning neutron stars near the Galactic Center, can identify ULDM solitons for scalar field masses \(m \lesssim 10^{-22}\,\mathrm{eV}\). These detectors also can constrain direct couplings of ULDM to the Standard Model, surpassing current limits and opening the previously loosely constrained region \(2 \times 10^{-22} \lesssim m \, [\mathrm{eV}] \lesssim 3 \times 10^{-21}\). A defining feature of this scenario is the universality of soliton-induced modulation effects, which are expected to influence all GW sources in the same region. This universality enables robust confirmation of the soliton's presence through the analysis of a population of GW sources.

In the main paper~\cite{Blas:2024duy}, we also investigated the case of ULDM with self-interactions and explored the detectability of solitons in extragalactic systems. Our findings suggest that GW modulation could probe solitonic structures at the centers of other galaxies, particularly in more massive dark matter halos. Additionally, we highlight the possibility of similar modulation effects arising at even lower frequencies due to the stochastic nature of dark matter halos. Investigating this phenomenon will be the focus of future work.

\section*{Acknowledgements}
This contribution is based on the paper~\cite{Blas:2024duy} with authors Diego Blas and Rodrigo Vicente. This article is based on the work from COST Action COSMIC WISPers CA21106, supported by COST (European Cooperation in
Science and Technology). I am grateful to the organizers of the \textit{2nd General Meeting of the COST Action COSMIC WSIPers} for the kind invitation to give this parallel talk. The author has the support of the \textit{AGAUR FI SDUR 2022} predoctoral program from the \textit{Department of Recerca i Universitats of Generalitat de Catalunya} and the European Social Plus Fund. The author also acknowledges the support from the \textit{Departament de Recerca i Universitats} from \textit{Generalitat de Catalunya} to the \textit{Grup de Recerca 00649 (Codi: 2021 SGR 00649)}.
The research leading to these results has received funding from the Spanish Ministry of Science and Innovation (PID2020-115845GB-I00/AEI/10.13039/501100011033) and acknowledges the support by the European Research Area (ERA) via the UNDARK project (project number 101159929).
IFAE is partially funded by the CERCA program of the Generalitat de Catalunya.


\FullConference{2nd Training School and General Meeting of the COST Action COSMIC WISPers  (CA21106) (COSMICWISPers2024)\\
 10-14 June 2024 and 3-6 September 2024\\
Ljubljana (Slovenia) and Istanbul (Turkey)\\}

\end{document}